\documentstyle[newarcrc,fleqn,psfig]{article}

\def\plotfig#1#2#3#4#5#6#7{\centering \leavevmode
    \vbox to#2{\rule{0pt}{#2}}
    \includegraphics{#1}}

\def\simlt{\lower.5ex\hbox{$\; \buildrel < \over \sim \;$}}
\def\simgt{\lower.5ex\hbox{$\; \buildrel > \over \sim \;$}}

\title { Spectral mapping of the accretion flow of UU Aquarii
		 \thanks{This work was partially supported by CNPq research grant
		 no.\ 300\,354/96-7 and by PRONEX grant FAURGS/FINEP 7697.1003.00.}}
\author{ R. Baptista \address{Departamento de F\'\i sica - CFM, UFSC,
		88040-900 Florian\'opolis, SC, Brazil, e-mail: bap@fsc.ufsc.br,
		silveira@fsc.ufsc.br}, C. Silveira $^{\rm a}$, J. Steiner
		\address{Laborat\'orio Nacional de Astrof\'\i sica-LNA/CNPq, CP 21,
		37500-000, Itajub\'a, Brazil, email: steiner@lna.br} and K. Horne
		\address{School of Physics \& Astronomy, University of St.\,Andrews,
		North Haugh, St.\,Andrews, Fife, KY16 9SS, Scotland, email:
		kdh1@st-and.ac.uk}}

\begin{document}
\maketitle

\begin{abstract}
Time-resolved spectroscopy of the novalike variable UU Aquarii is analyzed
with eclipse mapping techniques to produce spatially resolved spectra
of its accretion disc and gas stream as a function of distance from disc
centre in the range 3600-7000 \AA.  The spectrum of inner disc shows a
blue continuum filled with deep, narrow absorption lines which transition
to emission with clear P~Cygni profiles at intermediate and large radii.
The spectrum of the uneclipsed component has strong H\,I and He\,I emission
lines and Balmer jump in emission and is explained as optically thin
emission from a vertically extended disc wind.
Most of the line emission probably arises from the wind.
The spatially-resolved spectra also suggest the existence of gas stream
penetration in UU~Aqr, which can be seen down to $R\simeq 0.2\; R_{L1}$.
\end{abstract}

\section {Introduction}

UU~Aquarii is a bright eclipsing novalike (P$_{\rm orb}= 3.9$ hr) showing
long-term brightness variations of 0.3 mags \cite{BSC94,Honey}. Broad-band
eclipse mapping indicate that its inner disc is optically thick, with an
inferred mass accretion rate of \.{M}$= 10^{-9}\; {M_\odot \; yr^{-1}}$
\cite{BSH96}. It was suggested to be an SW Sex star based on its
relative flat radial temperature profile in the inner disc, 
single peaked asymmetric emission lines showing little eclipse,
large phase shift between photometric and spectroscopic conjunction
and orbital phase-dependent absorption in the Balmer lines \cite{BSH96,
Hoard}.

In this paper we report on the analysis of time-resolved spectroscopy
of UU Aqr with eclipse mapping techniques \cite{Horne} to derive
spatially-resolved spectra of the accretion flow in this binary.

\section {Observations and analysis}

Time-resolved optical spectroscopy ($\lambda\lambda$ 3500--6900 \AA) covering
5 eclipses of UU~Aqr was obtained with the 2.1-m telescope at KPNO on
July-August 1993 at a time resolution of 50\,s.
The observations were performed while UU Aqr was on its high brightness state.
The spectra were divided into 226 passbands (15 \AA\ in the continuum and
fainter lines, and $\simeq 500\; km\,s^{-1}$ across the most prominent lines)
and light curves were extracted for each passband. Average lightcurves for
each passband were obtained by combining the 5 individual lightcurves.
The eclipse mapping method was used to obtain a map of the disc brightness
distribution and the flux of an additional uneclipsed component in each
passband. The monochromatic maps were combined to produce spatially-resolved
spectra.

\section {Results and discussion}

Fig.\,1 shows spatially-resolved spectra of the disc region in a logarithmic
scale. The lower panel shows the spectrum of the uneclipsed light.
%
%
\begin{figure}[t]
\plotfig{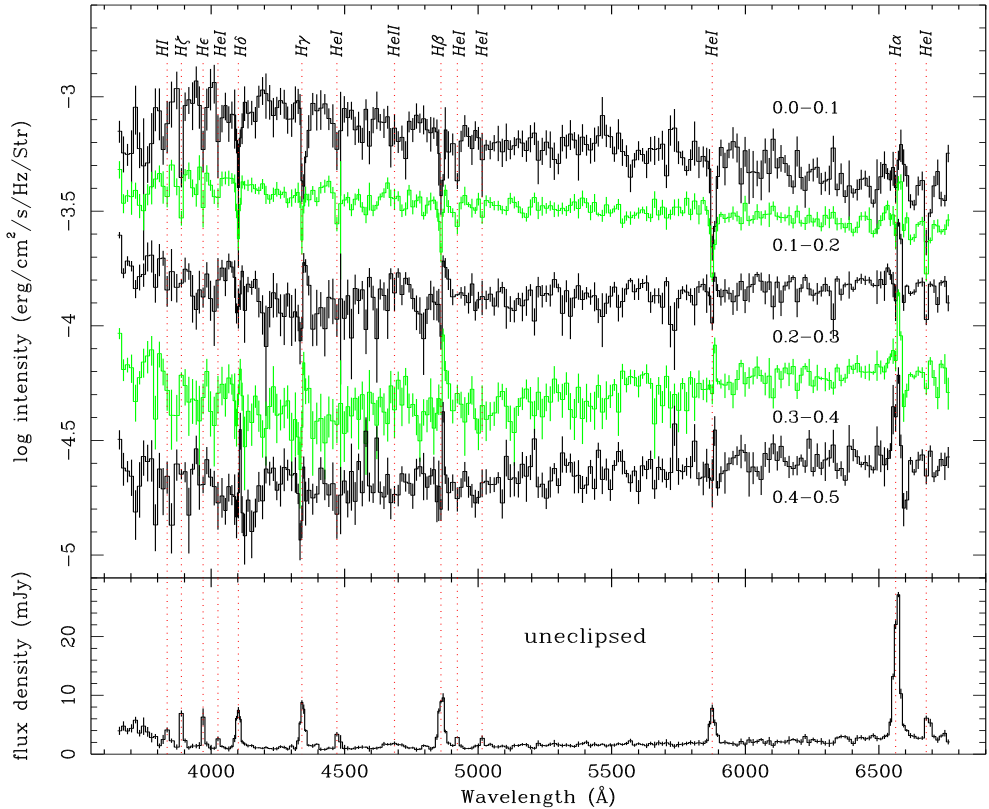}{9.6cm}{-90}{65}{63}{-255}{325}
\caption { Spatially resolved spectra of the UU Aqr accretion disc.
	The spectra were computed for a set of concentric annular sections
	(radius range indicated on the right, in units of $R_{L1}$) not 
	including the gas stream region. The inner annular region is at the
	top and each spectrum is at its true intensity level. 
	The lower panel shows the spectrum of the uneclipsed light.
	The most prominent line transitions are indicated by vertical dotted
	lines.  Error bars were derived via Monte Carlo simulations with the
	eclipse lightcurves. }
\end{figure}
The spectrum of the inner disc is characterized by a blue and bright
continuum filled with deep and narrow absorption lines. The continuum
emission becomes progressively fainter and redder for increasing disc radius
while the lines transition from absorption to emission showing clear P~Cygni
profiles on all lines mapped at higher spectral resolution. The Balmer jump
appears in absorption in the inner disc and weakly in emission in the
intermediate and outer disc regions, suggesting that the outer disc in
UU~Aqr is optically thin. The change in the slope and intensity of the
continuum with increasing disc radius reflects the temperature gradient in
the accretion disc, with the effective temperature decreasing outwards. 

Spatially resolved disc spectra are shown in Fig.\,2 as a function of
velocity for the H$\alpha$, H$\beta$ and H$\gamma$ lines.
The absorption lines at disc center are perceptibly narrower and deeper than
expected for emission from either the white dwarf atmosphere or from disc
gas in Keplerian orbits around the white dwarf, and the lines show larger
widths at the outer disc.
This is in clear disagreement with the expected behaviour of line emission
from gas in a Keplerian disc and is an evidence that most of
the line emission do not arise from the disc atmosphere.
On the other hand, the lines at intermediate and outer disc regions
($R \simgt 0.2\; R_{\rm L1}$) show clear P~Cygni profiles indicating origin
in an outflowing gas, probably the disc wind.
%
%
\begin{figure}
\plotfig{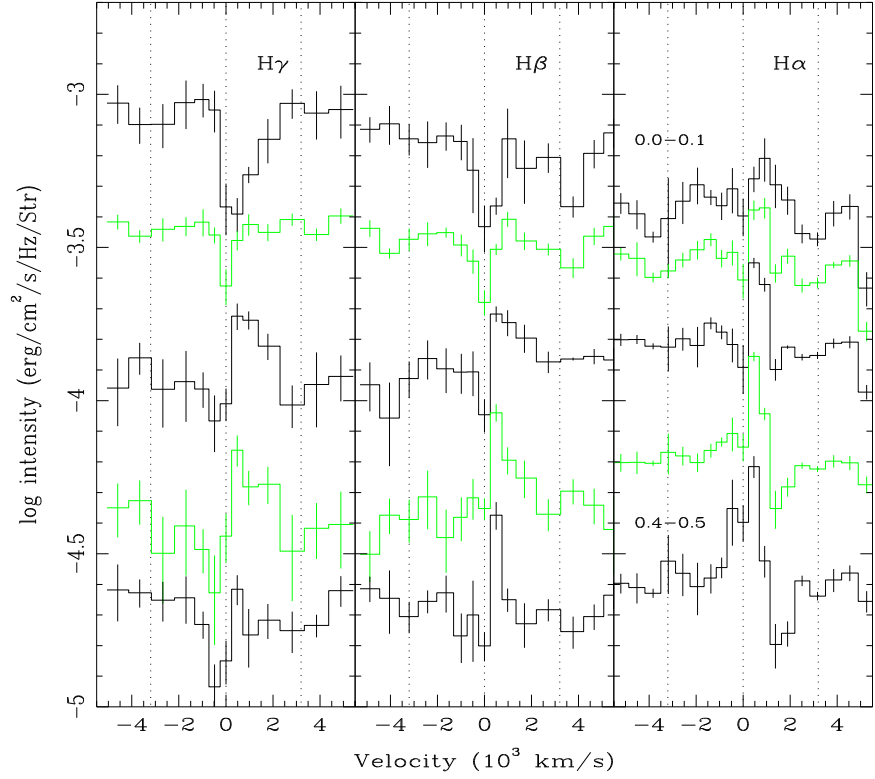}{9.2cm}{0}{65}{50}{-200}{-80}
\caption { Spatially resolved spectra in the H$\alpha$, H$\beta$ and
	H$\gamma$ regions as a function of velocity. The notation in the same
	as in fig.\,1. Dotted vertical lines mark line center and the maximum
	blueshift/redshift velocity expected for gas in Keplerian orbits around
	a $0.67 M_\odot$ white dwarf seen at an inclination of $i=78^o$
	($v \sin i= 3200 \;km \;s^{-1}$). }
\end{figure}

The spectrum of the uneclipsed light shows prominent Balmer and He\,I
emission lines (Fig.\,1). The Balmer jump (and possibly also the Paschen
jump) is clearly in emission indicating that the uneclipsed light has an
important contribution from optically thin gas from outside the orbital plane.
The fractional contribution of the uneclipsed light is very significant
for the emission lines, reaching 40-60\% at the Balmer lines and
20-40\% at the He\,I lines, and decreases steadily along the Balmer serie. 
The difference in fractional contribution between the Balmer and He\,I
lines indicates the existence of a vertical temperature gradient in the
material above/below the disc, with the He\,I lines (which require higher excitation energies) being produced closer to the orbital plane. In all
case, a substantial fraction of the light at these lines does not arise
from the orbital plane and is not occulted during eclipse.

Fig.\,3 shows the ratio between the spectrum of the gas stream region and
the disc region at same radius as a function of radius. 
The comparison reveals that the spectrum of the gas stream is noticeably
different from the disc spectrum in the outer disc regions (where one
expects a bright spot to form due to the initial shock between the inflowing
stream and the outer disc rim) but is also different in the inner disc
regions. This result suggests the existence of gas stream penetration
in UU~Aqr, which can be seen down to $R\simeq 0.2\; R_{L1}$.
The spectrum of the ratio becomes redder for decreasing disc radius,
possibly a combination of the disc emission becoming bluer as one moves
inwards and the gas stream emission becoming redder while its energy is
continuously lost in the shock with disc material along the inward stream
trajectory.
%
%
\begin{figure}
\plotfig{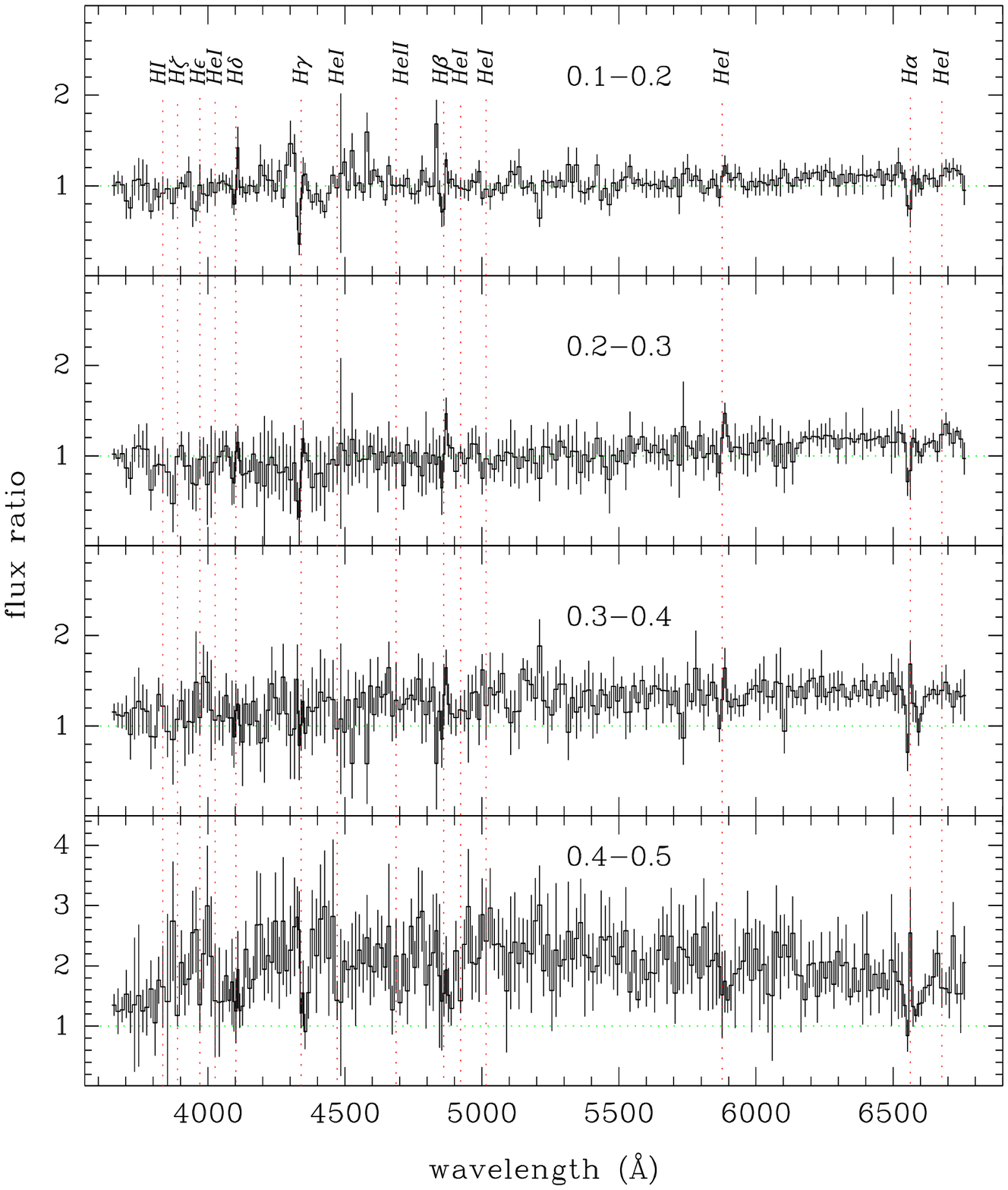}{9.8cm}{0}{60}{54}{-185}{-90}
\caption { Ratio of the gas stream spectrum and the disc spectrum at the
	same radius. A dotted line mark the unity level for each panel. The
	notation in the same as in fig.\,1. }
\end{figure}


\begin{thebibliography}{ Honeycutt et al, 1998 }

\bibitem [Baptista et al, 1994]{BSC94}
Baptista R., Steiner J. E., Cieslinski D., 1994, ApJ, 433, 332

\bibitem [Baptista et al, 1996]{BSH96}
Baptista R., Steiner J. E., Horne K., 1996, MNRAS, 282, 99

\bibitem [Hoard et al, 1998]{Hoard}
Hoard D. W., et~al., 1998, MNRAS, 294, 689

\bibitem [Honeycutt et al, 1998]{Honey}
Honeycutt R. K., Robertson J. W., Turner G. W., 1998, AJ, 115, 2527

\bibitem [Horne 1985]{Horne}
Horne, K. 1985, MNRAS, 213, 129

\end{thebibliography}
\end{document}